\documentclass[runningheads]{llncs}
\usepackage[T1]{fontenc}
\usepackage{graphicx,verbatim}
\usepackage[table]{xcolor}
\usepackage{hyperref}
\usepackage{color}

\usepackage{tikz}
\usepackage{siunitx}
\DeclareSIUnit\GB{GB}
\DeclareSIUnit\GiB{GiB}
\usepackage{amsmath}
\usepackage{graphicx}
\definecolor{pastelgreen}{RGB}{110, 210, 110}
\definecolor{pastelyellow}{RGB}{255, 240, 150}

\newcommand*\samethanks[1][\value{footnote}]{\footnotemark[#1]}

\begin{document}
\title{\emph{fastWDM3D}: Fast and Accurate 3D Healthy Tissue Inpainting}

\author{Alicia Durrer\inst{1}\orcidID{0009-0007-8970-909X} \and
Florentin Bieder\inst{1}\orcidID{0000-0001-9558-0623} \and
Paul Friedrich\inst{1}\orcidID{0000-0003-3653-5624} \and
Bjoern Menze\inst{2}\orcidID{0000-0003-4136-5690} \and
Philippe C. Cattin\inst{1}\orcidID{0000-0001-8785-2713}\thanks{equal contribution} \and
Florian Kofler\inst{2,3,4,5}\orcidID{0000-0003-0642-7884}\samethanks}


\authorrunning{A. Durrer et al.}
%
\institute{Department of Biomedical Engineering, University of Basel, Switzerland \and
Department of Quantitative Biomedicine, University of Zurich, Switzerland \and
Helmholtz AI, Helmholtz Zentrum München, Germany \and
Department of Diagnostic and Interventional Neuroradiology, School of Medicine, Klinikum rechts der Isar, Technical University of Munich, Germany \and
TranslaTUM - Central Institute for Translational Cancer Research, Technical University of Munich, Germany}

\maketitle              
\begin{abstract}
Healthy tissue inpainting has significant applications, including the generation of pseudo-healthy baselines for tumor growth models and the facilitation of image registration. In previous editions of the \emph{BraTS Local Synthesis of Healthy Brain Tissue via Inpainting Challenge}, denoising diffusion probabilistic models (DDPMs) demonstrated qualitatively convincing results but suffered from low sampling speed. To mitigate this limitation, we adapted a 2D image generation approach, combining DDPMs with generative adversarial networks (GANs) and employing a variance-preserving noise schedule, for the task of 3D inpainting.
Our experiments showed that the variance-preserving noise schedule and the selected reconstruction losses can be effectively utilized for high-quality 3D inpainting in a few time steps without requiring adversarial training.
We applied our findings to a different architecture, a 3D wavelet diffusion model (\emph{WDM3D}) that does not include a GAN component. The resulting model, denoted as \emph{fastWDM3D}, obtained a SSIM of $0.8571$, a MSE of $0.0079$, and a PSNR of $22.26$ on the \emph{BraTS} inpainting test set. Remarkably, it achieved these scores using only two time steps, completing the 3D inpainting process in \SI{1.81}{\second} per image.
When compared to other DDPMs used for healthy brain tissue inpainting, our model is up to $\sim  800 \times$ faster while still achieving superior performance metrics. Our proposed method, \emph{fastWDM3D}, represents a promising approach for fast and accurate healthy tissue inpainting. Our code is available at \url{https://github.com/AliciaDurrer/fastWDM3D}.


\end{abstract}

\section{Introduction}
Tumor growth datasets often lack images of initially healthy brains, which would be essential for accurately predicting tumor progression \cite{ezhov2023learn}. Furthermore, many automated brain magnetic resonance imaging (MRI) analysis tools, e.g., for segmentation, are trained exclusively on healthy data and perform better when pathological tissue is replaced with healthy tissue \cite{dadar2021beware}. Therefore, healthy tissue inpainting is a crucial step in both scenarios. The \emph{Brain Tumor Segmentation (BraTS)} challenge \cite{baid2021rsna,bakas2017advancing,menze2014multimodal} features an inpainting sub-challenge, highlighting the importance of this task. Inpainting challenge results \cite{durrer2023denoising,huo2023unleash,kofler2023brain,zhang2023synthesis} and a subsequent study \cite{durrer2024denoising}, comparing different denoising diffusion probabilistic models (DDPMs) \cite{ho2020denoising} on the \emph{BraTS} inpainting test set, show that methods using DDPMs yield promising outcomes. However, a major drawback is their long inference time. In this work, we present a modification of an inpainting-specific 3D wavelet diffusion model (\emph{WDM3D}) \cite{durrer2024denoising,friedrich2024wdm}. Our modification, called \emph{fastWDM3D}, only requires two time steps compared to the original 1000, but still manages to achieve better scores on the \emph{BraTS} inpainting test set.
\subsection{Related Work}
DDPMs \cite{ho2020denoising} produce diverse, high-quality outputs but require long sampling times, whereas generative adversarial networks (GANs) \cite{goodfellow2014generative} quickly generate high-quality samples but often face mode collapse. In \cite{xiao2021tackling}, Xiao et al. summarize this as the \textit{Generative Trilemma} and suggest modeling each denoising step in a diffusion model with multimodal conditional GANs (\emph{DDGANs}) to generate high-quality, diverse outputs fast. By using \emph{DDGANs} on wavelet transformed images (\emph{WDDGAN}), Phung et al. \cite{phung2023wavelet} further increase inference speed. Speeding up DDPMs is a recurring research topic: Reducing the number of time steps in training and inference has also been explored by e.g.,  \cite{frans2024one,karras2022elucidating,ye2024schedule}. 
Furthermore, distillation \cite{salimans2022progressive} or an optimized choice of time steps \cite{li2023autodiffusion,zheng2024beta} have been used to decrease inference time. Finding an efficient variance schedule, controlling the noise perturbation process, has been the focus of several works \cite{kingma2021variational,lin2024common,san2021noise,song2020score}. 
\subsection{Contribution}
In this study, we adapted the \emph{WDDGAN} \cite{phung2023wavelet}, originally designed for 2D natural image generation, for a 3D inpainting task. Our experiments revealed that the variance-preserving schedule \cite{song2020score} used in \emph{WDDGAN} is crucial for performance, while the adversarial training does not yield significant benefits for our application. Consequently, we eliminated the discriminator from the architecture, not only simplifying the model but also accelerating the training process without compromising the quality of the generated outputs.
To further substantiate the significance of the variance-preserving schedule and the reconstruction losses used, we integrated these components into \emph{WDM3D} [6,9], a DDPM-only architecture without any GAN component. The resulting model, termed \emph{fastWDM3D}, demonstrated superior performance compared to other DDPMs \cite{durrer2024denoising} evaluated on the \emph{BraTS} inpainting test set. Notably, it outperformed the original \emph{WDM3D} model, which employs 1000 time steps, a linear variance schedule, and a mean squared error (MSE) loss applied to wavelet coefficients \cite{durrer2024denoising,friedrich2024wdm}. In general, our \emph{fastWDM3D} achieves better performance metrics than all other assessed DDPMs, while preserving 3D consistency and being up to $\sim  800 \times$ faster.
\section{Methods}
\begin{figure}
\centering
\includegraphics[width=\textwidth]{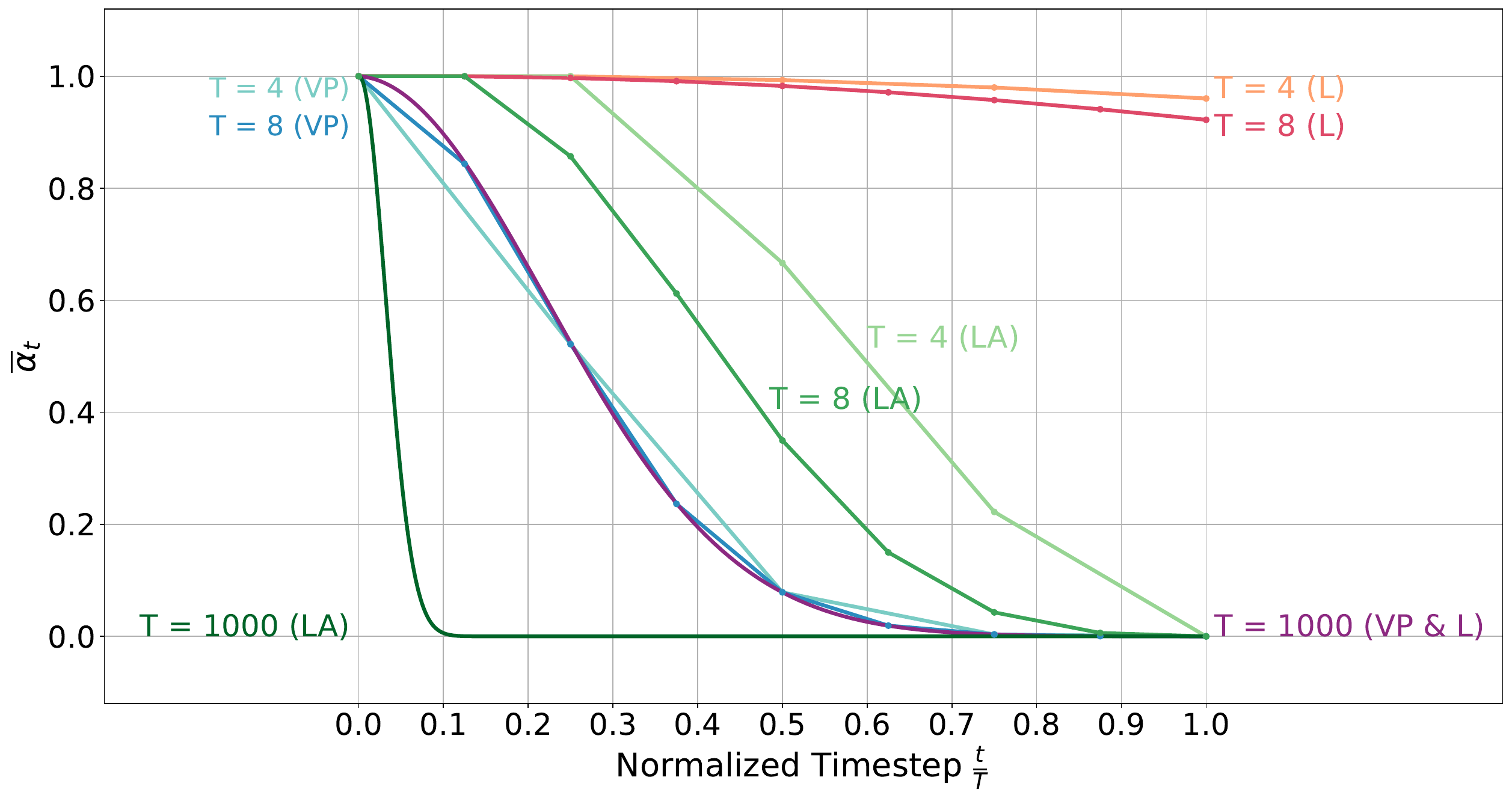}
\caption{Comparison of $\overline{\alpha}_{t}$ for the L-, LA- and VP schedule after normalizing all individual $T$ to [0,1]. The L- and VP schedules have the same curve for $T = 1000$. The L schedule only provides full perturbation for large $T$ while the LA schedule perturbs the image too early if $T$ is large. The VP schedule is applicable for low and large $T$.} \label{fig1}
\end{figure}
\subsection{Denoising Diffusion Probabilistic Models}
In DDPMs \cite{ho2020denoising}, noise gradually perturbs an input image $x_0$ for $T$ time steps:
\begin{equation}
q(x_{t}|x_{t-1}):=\mathcal{N}(x_{t};\sqrt{1-\beta _{t}}x_{t-1},\beta _{t}\vec{I}),
\label{eqn1}
\end{equation}
with $\vec{I}$ being the identity matrix and $\beta_{t}$ being the variance at time step $t$. This is called the \emph{forward process}.
The \emph{reverse} or \emph{denoising process}
\begin{equation}
p_{\theta}(x_{t-1}|x_t) := \mathcal{N}\left(x_{t-1}; \mu_{\theta}(x_t, t), \sigma_t^2\vec{I}\right)
\label{eqn2}
\end{equation}
follows a sequence of Gaussian distributions with mean $\mu_{\theta}$ and variance $\sigma_t^2$, parameterized by a time-conditioned model $\epsilon_{\theta }(x_{t},t)$. The goal of the reverse process is to match the true denoising distribution $q(x_{t-1}|x_{t})$.
The model $\epsilon_{\theta }(x_{t},t)$ can be trained to predict $x_0$, which then can be used to sample $x_{t-1}$ using the posterior distribution $q(x_{t-1}|x_{t},x_{0})$.
\subsection{Discrete Wavelet Transform}
Using generative models on the wavelet coefficients of the input images has been described in several works \cite{friedrich2024wdm,gal2021swagan,guth2022wavelet,phung2023wavelet}. The discrete wavelet transform (DWT) applies several low- and high-pass filters with stride 2, $l=\frac{1}{\sqrt{2}} \begin{bmatrix}1 & 1\end{bmatrix}$ and $h=\frac{1}{\sqrt{2}} \begin{bmatrix}-1 & 1\end{bmatrix}$, along all spatial dimensions. A 3D volume $y$ is decomposed into 8 wavelet coefficients ($x_{lll},x_{llh},x_{lhl},x_{lhh},x_{hll},x_{hlh},x_{hhl},x_{hhh}$) of half the spatial resolution of $y$. 
The coefficients can be concatenated into a single matrix $x$. The inverse discrete wavelet transform (IDWT) of $x$ restores the original image $y$.
\subsection{Variance-Preserving Noise Schedule}
Song et al. \cite{song2020score} showed that the noise perturbations in DDPMs, described by Eq.~\ref{eqn1}, correspond to a discretization of the stochastic differential equation (SDE)
$dx = -\frac{1}{2}\beta_{t}x \, dt + \sqrt{\beta_{t}} \, dw$,
where $ w $ is the standard Wiener process. This SDE is called Variance-Preserving (VP), as it leads to a process with bounded variance as $ t \to \infty $. Xiao et al. \cite{xiao2021tackling} rely on this SDE to compute the VP variance schedule
\begin{equation}
\beta_{t} = 1 - \exp\Big(-{\beta}_{\text{min}} \frac{t}{T} - 0.5({\beta}_{\text{max}} - {\beta}_{\text{min}}) \frac{2t-1}{T^2}\Big) \quad \text{for } t = 1,2,\ldots, T,
\label{eqn3}
\end{equation}
ensuring that the overall noise perturbation is independent of the number of diffusion steps, allowing full perturbation even if $T$ is low.
As $\beta_t$ determines the amount of noise added at each $t$, $\alpha _{t}:=1-\beta _{t}$ represents the original signal preservation at each $t$ and $\overline{\alpha}_{t}:=\prod_{s=1}^t \alpha _{s}$ accumulates the original signal preservation up until this $t$. Thus, the closer $\overline{\alpha}_{t}$ is to zero, the more the image is perturbed. The variance schedule defines the extent and speed of perturbation in the \emph{forward process}, whereby full perturbation is required but should not happen too early. Figure~\ref{fig1} shows the comparison of the linear variance schedule L (with $\beta_{1} = 10^{-4}$ and $\beta_T = 0.02$ as in \cite{ho2020denoising}), and the VP schedule (Eqn.~\ref{eqn3} with ${\beta}_{\text{min}} = 0.1$ and ${\beta}_{\text{max}} = 20$ to match the L schedule for $T=1000$ \cite{song2020score}). The VP schedule allows efficient noise perturbation using small and big $T$. In contrast, the L schedule does not perturb the image if $T$ is small. Scaling the L schedule by multiplying $\beta_{1}$ and $\beta_T$ by $\frac{1000}{T}$ is not possible, as $\beta>1$ if $T<20$. Thus, we also provide an adapted linear schedule (LA), with $\beta_{1} = 10^{-4}$ and $\beta_{T} = 0.9999$. We chose this $\beta_T$ to ensure that the image is fully perturbed in the \emph{forward process} when using small $T$. However, this is not applicable for large $T$, shown in Fig. \ref{fig1} for $T=1000$, as it destroys all image information in early stages already.
\subsection{Denoising Diffusion Generative Adversarial Networks}
In \emph{DDGAN}, the denoising distribution of a DDPM is modeled with an expressive multimodal distribution using a GAN. Fake samples from $p_{\theta}(x_{t-1}|x_t)$ are evaluated against real samples from $q(x_{t-1}|x_t)$. An adversarial loss, $\mathcal{L}_{adv}$, minimizes the softened reverse KL divergence per denoising step.
The generator is based on the NCSN++ architecture \cite{song2020score} and is conditioned on a latent variable provided to the NCSN++ using a mapping network, originating from StyleGAN \cite{karras2019style}. 
\emph{WDDGAN} further reduces sampling times by working in the wavelet domain.
\subsection{Models Used for Healthy Tissue Inpainting}
We employed three distinct network architectures, listed below, that we adapted for inpainting through \emph{Palette}-conditioning \cite{saharia2022palette}. For the training process, we used the ground truth image ($g$), a mask masking out some healthy tissue ($m$), and the resulting voided image ($v$). The wavelet coefficients of $v$, $m$, and the noisy $g$ were concatenated to form a 24-channel input. This input was then processed by the respective model, generating an 8-channel output, which was subsequently subjected to the IDWT to reconstruct the inpainted prediction, denoted as $\hat{y}$.
\begin{itemize}
    \item \emph{WDDGAN3D}: We extended the \emph{WDDGAN} to a 3D inpainting network. In addition to the adversarial loss $\mathcal{L}_{adv}$ we incorporated a reconstruction loss $\mathcal{L}_{{\hat{y}}}$ between $g$ and $\hat{y}$. Additionally, we computed a region-specific reconstruction loss $\mathcal{L}_{{\hat{y}_m}}$ focused on the masked area $m$ of both $g$ and $\hat{y}$. The total loss $\mathcal{L}$ was defined as $\mathcal{L} = \mathcal{L}_{adv} + \mathcal{L}_{{\hat{y}}} + \mathcal{L}_{{\hat{y}_m}}$. We assessed the performance for $T \in \{4, 8\}$ in different setups, including adjustments to the frequency of generator updates relative to the discriminator and variations in learning rates. However, all of them led to the same outcome: $\mathcal{L}_{adv}$ did not decrease during training. We report the vanilla configuration in Table~\ref{tab1}.
    \item \emph{GO3D}: Given that $\mathcal{L}_{adv}$ did not decrease during training, we removed the discriminator and the associated $\mathcal{L}_{adv}$, resulting in a simplified loss function defined as $\mathcal{L} = \mathcal{L}_{{\hat{y}}} + \mathcal{L}_{{\hat{y}_m}}$. This model is referred to as \emph{Generator-only 3D} (\emph{GO3D}) and we evaluted its performance for $T \in \{2, 4, 8, 16, 64, 256, 1000\}$.
    \item \emph{fastWDM3D}: Building on the work of Durrer et al. \cite{durrer2024denoising}, who demonstrated the efficacy of the wavelet diffusion model 3D (\emph{WDM3D}) by Friedrich et al. \cite{friedrich2024wdm} for inpainting tasks, we modified the noise schedule of the \emph{WDM3D} by implementing the variance-preserving (VP) schedule. Additionally, we replaced the MSE loss on the wavelet coefficients with the loss function $\mathcal{L} = \mathcal{L}_{{\hat{y}}} + \mathcal{L}_{{\hat{y}_m}}$. Our training utilized time steps $T \in \{2, 4, 8\}$, contrasting with the $T = 1000$ employed in \cite{durrer2024denoising}. This model is termed \emph{fastWDM3D}.
\end{itemize}
\section{Experimental Details and Results}
\subsubsection{Dataset}
We used the publicly available dataset of the
\emph{BraTS 2023 Local Synthesis of Healthy Brain Tissue via Inpainting Challenge} \cite{baid2021rsna,bakas2017advancing,karargyris2023federated,menze2014multimodal} for training. It contains T1 scans showing brain tumors, as well as masks of the tumor and masks of some regions showing only healthy tissue (called healthy masks). From the 1251 patient scans, we used 1200 to train our models and 51 for validation to observe model performance during training. The non-public \emph{BraTS} inpainting test set contains T1 scans, healthy masks and tumor masks of 568 patients. All scans had an initial resolution of $240\times 240\times 155$. We removed the top and bottom 0.5 percentile of voxel intensities and normalized them between -1 and 1. For training, we cropped out the region defined by the healthy mask $m$ in the T1 image $g$, generating a voided image $v$. The images $v$, $m$ and $g$ were all cropped to $128\times 128\times 128$ with the region to be inpainted in the center. Each prediction $\hat{y}$ was normalized back to the individual input range of $v$ for evaluation.
\subsubsection{Experiments and Results}
All models were trained on a NVIDIA A100 (\SI{40}{\GB}) GPU. All implementation details can be found at \url{https://github.com/AliciaDurrer/fastWDM3D}.
We first compared \emph{WDDGAN3D} and \emph{GO3D}, see Table~\ref{tab1}. For \emph{GO3D} we explored the LA and the VP schedule, for \emph{WDDGAN3D} we used the VP schedule only. We evaluated the models after 100 epochs, as we could observe convergence followed by overfitting for all model types. We report training duration and memory requirement, as well as SSIM, MSE and PSNR on the \emph{BraTS} inpainting test set. \emph{GO3D(LA)} showed the best performance regarding MSE, however, it was worse than \emph{WDDGAN3D} and \emph{GO3D} regarding SSIM and PSNR. As \emph{GO3D(VP)} performed better than \emph{WDDGAN3D} across all metrics while requiring less memory and significantly less training time, we evaluated \emph{GO3D(VP)} for further time steps. An additional reason for keeping the VP instead of the LA schedule was that the VP schedule can be used independent of the number of time steps $T$ without requiring any modifications (see Fig.\ref{fig1}). Since our goal is to keep $T$ as low possible, this is not decisive, but we want to provide a robust method that allows changes if required.
\begin{table}
\centering
\caption{Comparison of \emph{WDDGAN3D} using the VP schedule, \emph{GO3D} using the LA schedule, and \emph{GO3D} using the VP schedule on the \emph{BraTS} inpainting test set. For all experiments: batch size: 2, residual blocks in generator: 4, learning rate generator: $2\cdot 10^{-5}$, trained for 100 epochs ($60\cdot10^{3}$ iterations). Abbreviations: T = time steps, TT = training time, Mem = memory required during training.}\label{tab1}
\begin{tabular}{c|c|c|c|c|c|c}
Type & T & TT ($\downarrow$) & Mem ($\downarrow$) & SSIM ($\uparrow$) & MSE ($\downarrow$) & PSNR $(\uparrow$) \\
\hline
\emph{WDDGAN3D(VP)} & 4 & $\sim$\SI{112}{\hour} & \SI{31.69}{\GiB} & $0.8562 \pm 0.1169$ &  $0.0082 \pm 0.0063$ &  $21.92 \pm 3.74$\\
\emph{WDDGAN3D(VP)} & 8 & $\sim$\SI{112}{\hour} & \SI{31.69}{\GiB} & $0.8587 \pm 0.1165$ & $0.0081 \pm 0.0064$ & $22.08 \pm 3.84$\\
\emph{GO3D(VP)} & 4 & $\sim$\SI{11}{\hour} & \SI{24.24}{\GiB} & $0.8595 \pm 0.1162$ & $0.0079 \pm 0.0062$ & $22.17 \pm 3.84$\\
\rowcolor{pastelyellow}
\emph{GO3D(VP)} & 8 & $\sim$\SI{11}{\hour} & \SI{24.24}{\GiB} & $0.8606 \pm 0.1145$ & $0.0079 \pm 0.0061$ & $22.19 \pm 3.83$\\
\emph{GO3D(LA)} & 4 & $\sim$\SI{11}{\hour} & \SI{24.24}{\GiB} & $ 0.8521 \pm 0.1242$ & $0.0066 \pm 0.0065$ & $21.69\pm 4.10$\\
\emph{GO3D(LA)} & 8 & $\sim$\SI{11}{\hour} & \SI{24.24}{\GiB} & $0.8438 \pm 0.1409$ & $0.0061 \pm 0.0049$ & $21.83 \pm 3.54$\\
\end{tabular}
\end{table}
\begin{table}
\centering
\caption{Comparison of \emph{GO3D} and \emph{fastWDM3D} on the \emph{BraTS} inpainting test set, both use the VP schedule. For all experiments: batch size:~3, residual blocks:~4, learning rate:~$2\cdot 10^{-5}$. Abbreviations: T = time steps, Iter = training iterations (as a multiple of $10^{3}$), AST = average sampling time per volume (can vary as it depends on the overall server utilization).}\label{tab2}
\begin{tabular}{c|r|r|c|c|c|c}
Type & T\hphantom{ii} & Iter & SSIM ($\uparrow$) & MSE ($\downarrow$) & PSNR ($\uparrow$) & AST ($\downarrow$) \\
\hline
\emph{GO3D} & 2 & 40 & $0.8553 \pm 0.1165$ & $0.0084 \pm 0.0064$ & $21.77 \pm 3.68$ & \hphantom{00}\SI{1.45}{\second} \\
\emph{GO3D} & 4 & 40 &  $0.8596 \pm 0.1147$ & $0.0080 \pm 0.0061$  & $22.03 \pm 3.69$ & \hphantom{00}\SI{2.37}{\second} \\
\emph{GO3D} & 8 & 40 &  $0.8596 \pm 0.1156$ & $0.0080\pm 0.0062$ & $22.08 \pm 3.74$ & \hphantom{00}\SI{4.25}{\second} \\
\emph{GO3D} & 16 & 40 & $0.8599 \pm 0.1155$ & $0.0079 \pm 0.0061$ & $22.11 \pm 3.73$ & \hphantom{00}\SI{7.95}{\second} \\
\emph{GO3D} & 64 & 40 & $0.8567 \pm 0.1175$ & $0.0083 \pm 0.0643$ & $21.93 \pm 3.77$ & \hphantom{0}\SI{62.87}{\second} \\
\emph{GO3D} & 256 & 40 & $0.8518 \pm 0.1200$ & $0.0090 \pm 0.0071$ & $21.65 \pm 3.91$ & \SI{118.46}{\second} \\
\emph{GO3D} & 1000 & 40 & $0.8483 \pm 0.1206$ &$0.0092 \pm 0.0070$ & $21.52 \pm 3.92$ & \SI{526.67}{\second} \\
\hline
\emph{GO3D} & 64 & 120 & $0.8496 \pm 0.1228$ & $0.0085 \pm 0.0067$ & $21.91 \pm 3.94$ & \hphantom{0}\SI{62.87}{\second} \\
\emph{GO3D} & 256 & 120 & $0.8502 \pm 0.1226$ & $0.0084 \pm 0.0066$ & $21.97 \pm 3.98$ & \SI{118.46}{\second} \\
\emph{GO3D} & 1000 & 120 &  $0.8507 \pm 0.1213$ & $0.0084 \pm 0.0066$ & $21.97 \pm 3.97$ & \SI{526.67}{\second} \\
\hline
\emph{fastWDM3D} & 2 & 40 & $0.8541 \pm 0.1177$ & $0.0090 \pm 0.0066$ & $21.42 \pm 3.56$ & \hphantom{00}\SI{1.81}{\second} \\
\emph{fastWDM3D} & 4 & 40 & $0.8366 \pm 0.1281$ & $0.0121 \pm 0.0091 $ & $20.17 \pm 3.70$ & \hphantom{00}\SI{3.78}{\second} \\
\emph{fastWDM3D} & 8 & 40 & $0.8072 \pm 0.1481$ & $0.0167 \pm 0.0135$ & $18.96 \pm 3.97$ & \hphantom{00}\SI{4.10}{\second} \\
\hline
\rowcolor{pastelgreen}
\emph{fastWDM3D} & 2 & 120 & $0.8571 \pm 0.1193$ & $0.0079 \pm 0.0063$ & $22.26 \pm 3.97$ & \hphantom{00}\SI{1.81}{\second} \\
\emph{fastWDM3D} & 4 & 120 & $0.8566 \pm 0.1185$ & $0.0081 \pm 0.0065$ & $22.20 \pm 4.03$ & \hphantom{00}\SI{3.78}{\second} \\
\emph{fastWDM3D} & 8 & 120 & $0.8561 \pm 0.1192$ & $0.0079 \pm 0.0063$ & $22.24 \pm 4.01$ & \hphantom{00}\SI{4.10}{\second} \\
\end{tabular}
\end{table}
\begin{table}
\centering
\caption{Comparison of the \emph{BraTS} 2023 inpainting challenge podium and our best model, \emph{fastWDM3D} ($T=2$), on the \emph{BraTS} inpainting test set.}\label{tab3}
\begin{tabular}{l|c|c|c}
 &  SSIM ($\uparrow$) & MSE ($\downarrow$) & PSNR ($\uparrow$) \\
\hline
Zhang et al. (1st 2023) \cite{zhang2023synthesis} & $0.91 \pm 0.15$ & $0.0049 \pm 0.0016$ & $23.59 \pm 5.35$ \\
Durrer et al. (2nd 2023) \cite{durrer2023denoising} & $0.86 \pm 0.20$ & $0.0100 \pm 0.0016$ & $20.42 \pm 3.82$ \\
Huo et al. (3rd 2023) \cite{huo2023unleash} & $0.87 \pm 0.18$ & $0.0144 \pm 0.0025$ & $18.71 \pm 4.01$ \\
Ours (\emph{fastWDM3D} ($T=2$)) & $0.86 \pm 0.12$ & $0.0079 \pm 0.0063$ & $22.26 \pm 3.97$ \\
\end{tabular}
\end{table}
We then compared \emph{GO3D(VP)} to \emph{fastWDM3D}, also using the VP schedule, in Table~\ref{tab2}. We report SSIM, MSE, PSNR, and average sampling time per volume on the \emph{BraTS} inpainting test set. For these experiments, the batch size was 3 instead of 2 used for the experiments in Table~\ref{tab1}. We report the scores obtained on the test set after training the models for 100 epochs ($40 \cdot 10^{3}$ iterations) as there was no further improvement of \emph{GO3D} on the validation set after these epochs. For \emph{fastWDM3D}, we additionally provide the results after 300 epochs ($120 \cdot 10^{3}$ iterations), as we saw that its performance on the validation set kept increasing after $100$ epochs. We also report the results for \emph{GO3D} after 300 epochs for $T \in \{64, 256, 1000\}$ to demonstrate that more training iterations do not improve performance compared to lower $T$ using less iterations. \emph{GO3D} requires \SI{38.8}{\GiB} and \emph{fastWDM3D} requires \SI{18.3}{\GiB} during training. 
\begin{figure}
    \centering
    \begin{tikzpicture}
        \node at (0, 1.4) {Voided};
        \node at (2, 1.4) {GT};
        \node at (4, 1.4) {$1^{st}$ 2023};
        \node at (6, 1.4) {$2^{nd}$ 2023};
        \node at (8, 1.4) {$3^{rd}$ 2023};
        \node at (10, 1.4) {\emph{Ours}};
        
        \node at (0, 0) {\includegraphics[width=0.15\textwidth, angle=180]{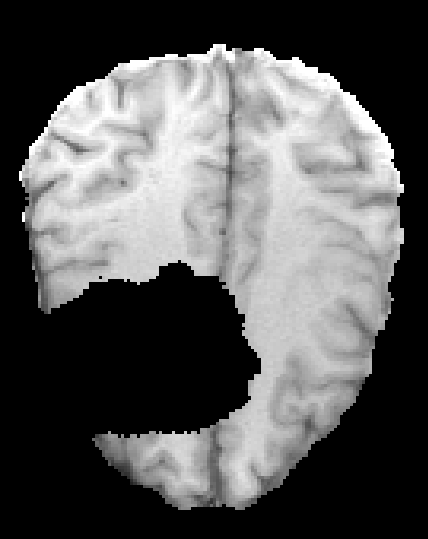}};
        \node at (2, 0) {\includegraphics[width=0.15\textwidth, angle=180]{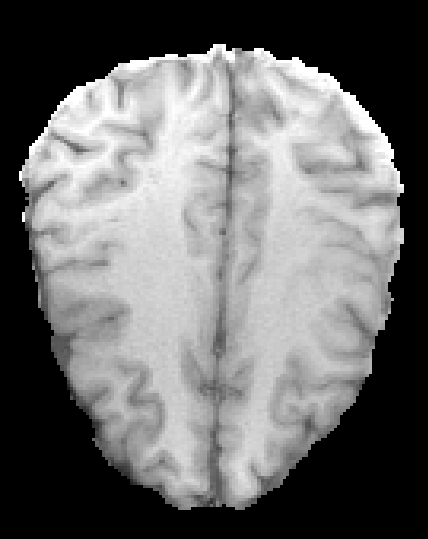}};
        \node at (4, 0) {\includegraphics[width=0.15\textwidth, angle=180]{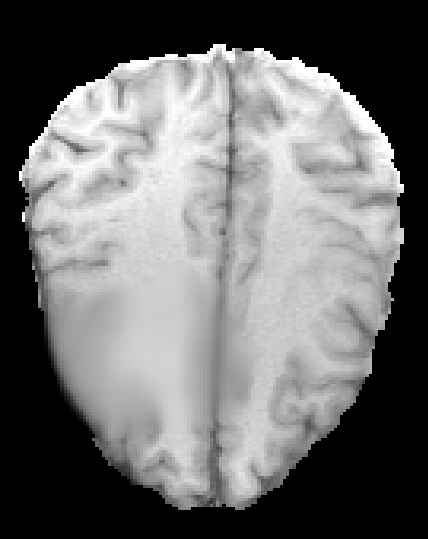}};
        \node at (6, 0) {\includegraphics[width=0.15\textwidth, angle=180]{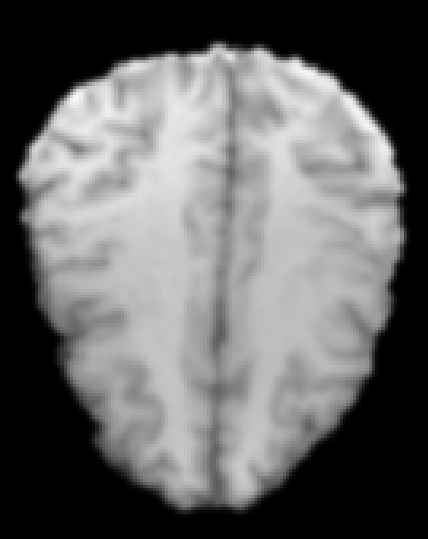}};
        \node at (8, 0) {\includegraphics[width=0.15\textwidth, angle=180]{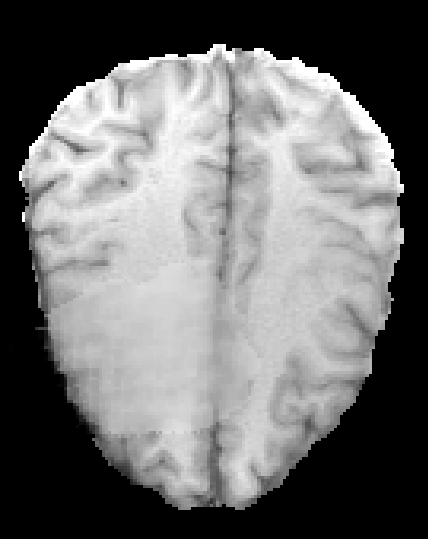}};
        \node at (10, 0) {\includegraphics[width=0.15\textwidth, angle=180]{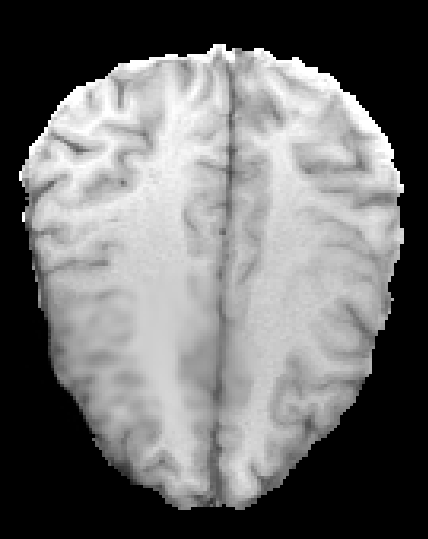}};
        
        \node at (0, -1.93) {\reflectbox{\includegraphics[width=0.15\textwidth]{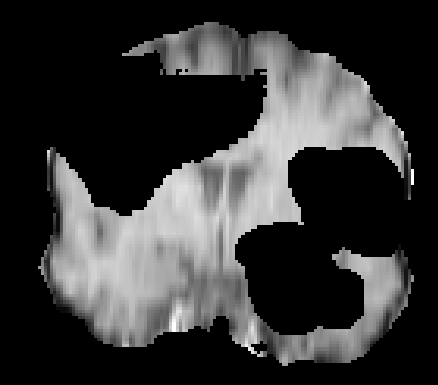}}};
        \node at (2, -1.93) {\reflectbox{\includegraphics[width=0.15\textwidth]{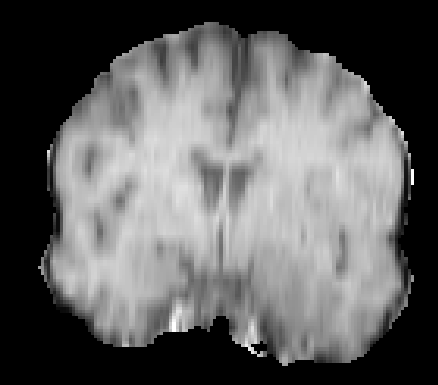}}};
        \node at (4, -1.93) {\reflectbox{\includegraphics[width=0.15\textwidth]{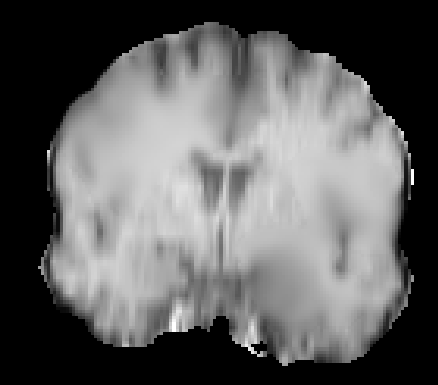}}};
        \node at (6, -1.93) {\reflectbox{\includegraphics[width=0.15\textwidth]{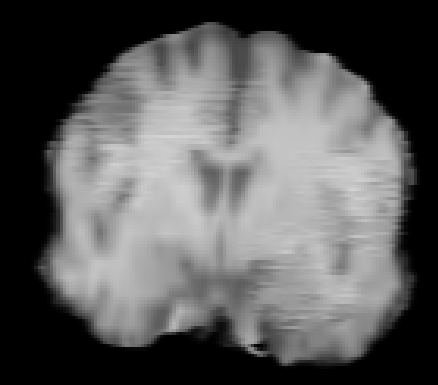}}};
        \node at (8, -1.93) {\reflectbox{\includegraphics[width=0.15\textwidth]{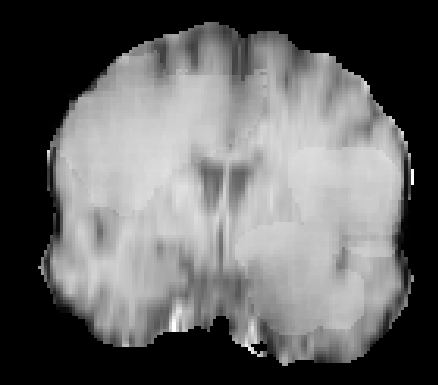}}};
        \node at (10, -1.93) {\reflectbox{\includegraphics[width=0.15\textwidth]{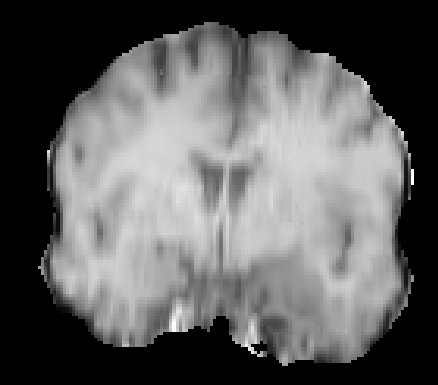}}};
    \end{tikzpicture}
    \caption{Axial (top) and coronal (bottom) view of an image of the validation set. Comparison of the \emph{BraTS} 2023 inpainting challenge podium and our best model, \emph{fastWDM3D} ($T=2$), given the voided input image (voided) and the ground truth (GT).}
\label{fig2}
\end{figure}
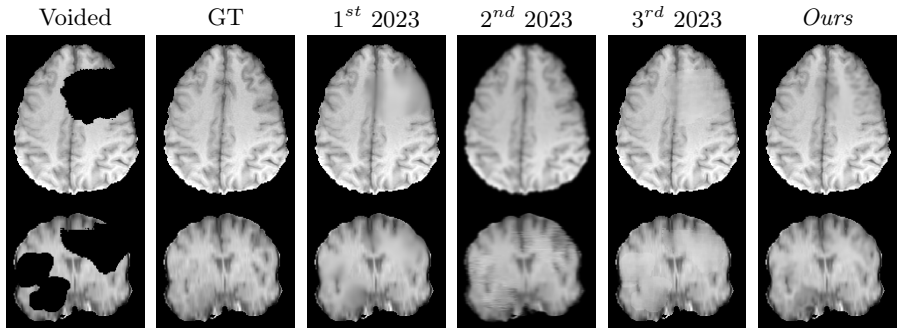
\begin{table}
\centering
\caption{Comparison of DDPM configurations by Durrer et al. \cite{durrer2024denoising} and our best model, \emph{fastWDM3D} ($T=2$), on the \emph{BraTS} inpainting test set. Abbreviation: AST = average sampling time per volume (can vary as it depends on the overall server utilization).}\label{tab4}
\begin{tabular}{l|c|c|c|c}
 &  SSIM ($\uparrow$) & MSE ($\downarrow$) & PSNR ($\uparrow$) & AST ($\downarrow$) \\
\hline
\emph{DDPM 2D slice-wise} \cite{durrer2024denoising} & $0.78 \pm 0.15$ & $0.0160 \pm 0.0118$ & $18.71 \pm 3.08$ & \SI{20}{\minute} \\
\emph{DDPM 2D seq-pos} \cite{durrer2024denoising} & $0.77 \pm 0.20$ & $0.0420 \pm 0.1136$ & $18.45 \pm 5.28$ & \SI{25}{\minute} \\
\emph{DDPM Pseudo3D} \cite{durrer2024denoising} & $0.85 \pm 0.12$ & $0.0103 \pm 0.0107$ & $20.93 \pm 3.38$ & \SI{25}{\minute} \\
\emph{DDPM 3D mem-eff} \cite{durrer2024denoising} & $0.70 \pm 0.16$ & $0.0523 \pm 0.0222$ & $14.14 \pm 3.18$ & \SI{20}{\minute} \\
\emph{LDM3D} \cite{durrer2024denoising} & $0.60 \pm 0.12$ & $0.0700 \pm 0.0280$ & $\hphantom{0}8.77 \pm 1.89$ & \hphantom{0}\SI{1}{\minute} \\
\emph{WDM3D} \cite{durrer2024denoising} & $0.61 \pm 0.16$ & $0.1060 \pm 0.0757$ & $10.57 \pm 3.20$ & \hphantom{0}\SI{5}{\minute} \\
Ours (\emph{fastWDM3D} ($T=2$)) & $0.86 \pm 0.12$ & $0.0079 \pm 0.0063$ & $22.26 \pm 3.97$ & \hphantom{0}\SI{1.81}{\second} \\
\end{tabular}
\end{table}
Based on Tables~\ref{tab1} and \ref{tab2}, we observe that \emph{GO3D} trained with a batch size of 2 and $T=8$ achieves the best SSIM, while \emph{fastWDM3D} trained for $120\cdot 10^3$ iterations with a batch of size 3 and $T=2$ obtains the best PSNR. The rounded MSE is the same for both. These methods are highlighted in yellow and green in Table \ref{tab1} and \ref{tab2}, respectively. As \emph{fastWDM3D} ($T=2$) additionally has the lowest average sampling time ($1.81 \text{s}$) and uses significantly less memory during training ($18.3$ GiB), we labeled this as our best configuration and compared it qualitatively and quantitatively to the \emph{BraTS} 2023 inpainting challenge podium \cite{durrer2023denoising,huo2023unleash,kofler2023brain,zhang2023synthesis}. The challenge proceedings of 2024 are not available yet. For the qualitative comparison, we sampled images of the 2023 challenge podium methods using the publicly available algorithms \cite{bratsalgorithms} and provide exemplary axial and coronal slices of the inpainting generated by all methods in Fig.~\ref{fig2}. The scores obtained on the \emph{BraTS} inpainting test set are reported in Table~\ref{tab3}. In a final step, we compared our best configuration to the different DDPMs by \cite{durrer2024denoising}, summarized in Table~\ref{tab4}. Among these configurations is their \emph{WDM3D} using a linear variance schedule, an MSE loss on the wavelet coefficients and $T=1000$.
\section{Discussion and Conclusion}
Upon analyzing the results in Tables~\ref{tab1} and \ref{tab2}, we see that the SSIM, MSE and PSNR are on a similar level for all evaluated model types (\emph{WDDGAN3D}, \emph{GO3D}, \emph{fastWDM3D}). However, the \emph{fastWDM3D} has further benefits: The configuration achieving the highest scores, \emph{fastWDM3D} ($T=2$), is also the fastest in sampling, requiring on average only \SI{1.81}{\second} for the full 3D inpainting. In addition, its slimmer architecture uses much less memory than \emph{WDDGAN3D} and \emph{GO3D}. 
Comparing our \emph{fastWDM3D} ($T=2$) to the podium of the \emph{BraTS 2023 Local Synthesis of Healthy Brain Tissue via Inpainting Challenge} \cite{kofler2023brain}, shown in Table~\ref{tab3}, we see that our method outperforms the methods achieving the second \cite{durrer2023denoising} and third \cite{huo2023unleash} place in terms of MSE and PSNR while being similar in terms of SSIM. The method placed first \cite{zhang2023synthesis} in the challenge achieves quantitatively better scores than our proposed method. However, the qualitative results presented in Fig.~\ref{fig2} show that our method inpaints 3D realistic structures, while the methods achieving the first and third place create more blurry and less defined inpainting. The second place, a 2D DDPM, generates realistic inpainting in the axial plane, but, the 2D nature of the model leads to stripe artifacts visible in the coronal view. 
Compared to the diffusion model setups by \cite{durrer2024denoising}, presented in Table~\ref{tab4}, our method achieves better SSIM, MSE and PSNR scores while being up to $\sim 800 \times$ faster during sampling. The key role of the VP schedule and the reconstruction losses is visible in Table~\ref{tab4} when comparing our scores to those of the \emph{WDM3D} inpainting implementation by \cite{durrer2024denoising}. Changing the schedule and the loss allowed a vast improvement in the scores while using $500 \times$ fewer time steps.

Our proposed method, \emph{fastWDM3D}, shows promising potential for clinical applications as it generates high-quality inpainting in a short amount of time. Fast training and sampling, as well as low memory requirements result in a smaller environmental footprint. Future research will include a closer investigation into why the VP schedule enables this performance and speed boost. Moreover, we aim to explore additional schedules, as the simply designed LA schedule already appears to be a promising approach. In addition, an ablation study disentangling the influence of the variance schedule, the loss, and the model architecture would be beneficial. Experimenting with different datasets and modalities, as well as exploring additional generative tasks (e.g., unconditional image generation), will further assess the robustness of our method.  

    

\begin{credits}

\subsubsection{\discintname}
The authors have no competing interests to declare that are relevant to the content of this article.
\end{credits}


%
%
%
\bibliographystyle{splncs04}
\bibliography{Paper-0029}
%




\end{document}